\documentclass[sigconf,screen,nonacm,review=false,timestamp=false]{acmart}

\AtBeginDocument{%
  }

\setcopyright{acmlicensed}
\copyrightyear{2026}
\acmYear{2026}
\acmDOI{XXXXXXX.XXXXXXX}
\acmConference[The ACM Web Conference]{The ACM Web Conference}{April 2026}{Dubai, United Arab Emirates}

\begin{document}

\title{Cooperation Under Network-Constrained Communication}


\author{Tommy Mordo}
\affiliation{%
  \institution{Technion}
  \city{Haifa}
  \country{Israel}}
\email{tommymordo@technion.ac.il}

\author{Omer Madmon}
\affiliation{%
  \institution{Technion}
  \city{Haifa}
  \country{Israel}}
\email{omermadmon@campus.technion.ac.il}

\author{Moshe Tennenholtz}
\affiliation{%
  \institution{Technion}
  \city{Haifa}
  \country{Israel}}
\email{moshet@technion.ac.il}

\renewcommand{\shortauthors}{Mordo et al.}


\begin{abstract}
In this paper, we study cooperation in distributed games under network-constrained communication. Building on the framework of Monderer and Tennenholtz (1999), we derive a sufficient condition for cooperative equilibrium in settings where communication between agents is delayed by the underlying network topology. Each player deploys an agent at every location, and local interactions follow a Prisoner’s Dilemma structure. We derive a sufficient condition that depends on the network diameter and the number of locations, and analyze extreme cases of instantaneous, delayed, and proportionally delayed communication. We also discuss the asymptotic case of scale-free communication networks, in which the network diameter grows sub-linearly in the number of locations. These insights clarify how communication latency and network design jointly determine the emergence of distributed cooperation.
\end{abstract}




\maketitle

\section{Introduction}

Distributed games, as introduced by \citet{monderer1999distributed}, provide a foundational framework for studying strategic behavior when each player deploys a set of local agents that interact across multiple locations. In such environments, coordination and cooperation may emerge even when local interactions are strictly competitive, provided that players’ agents can effectively share information about others’ behavior. However, real-world distributed systems rarely allow instantaneous communication: information typically propagates through a network, often with latency, congestion, or failure. These communication delays can affect agents’ ability to detect deviations and sustain cooperative equilibria.\footnote{Such communication delays can be viewed as inducing an endogenous form of imperfect recall, conceptually related to the framework of \citet{tewolde2024imperfect}.
}

In this paper, we extend the distributed-games framework to \emph{network-constrained communication}. We assume that agents are positioned at nodes of a communication network, where messages can be exchanged only along existing edges and reach other locations after a delay determined by the network topology. This restriction introduces a new dimension of strategic uncertainty: a deviation at one location may remain undetected elsewhere until several rounds later, potentially undermining cooperative behavior. Our goal is to characterize the conditions under which full cooperation can still arise in equilibrium when communication is delayed or localized.

We focus on the canonical case of two players who each deploy an agent at every network location. The agents play a \emph{Prisoner’s Dilemma (PD)} \cite{axelrod1980effective,axelrod1981emergence,axelrod1987evolution,rapoport2018prisoner} at each site, reflecting the tension between individual incentives to defect and collective incentives to cooperate. The order in which locations become active is uniformly random. We derive a sufficient condition for global cooperation in environments with network delays, and analyze its dependence on the number of locations and on the topological properties of the communication network (in particular, its diameter).

As a motivating example, consider two security applications, each controlled by a different vendor, that are deployed across a network of servers. At every server, the two applications can either \emph{share threat-detection data (cooperate)} or \emph{withhold it (defect)}. Sharing improves the overall detection rate for both but incurs computation and communication costs, while withholding data saves local resources but may compromise the joint detection capability. Locally, this interaction is a \emph{Prisoner’s Dilemma}: each application prefers to defect if the other cooperates, yet both are better off if cooperation occurs at all servers.\footnote{A related game-theoretic security model was introduced by \citet{oren2013pay}, where \emph{cooperation} corresponds to investing in antivirus protection, and \emph{defection} to free-riding on others to block the spread of infection.}

To coordinate cooperation, applications rely on their distributed agents to exchange alerts across the network. If one agent defects (e.g., withholds data), other agents can issue ``alarm'' messages to signal deviation, but these alarms take time to propagate due to network latency. The resulting \emph{delay in detectability} determines how long a deviation remains profitable before being punished elsewhere. Understanding how network structure affects the sustainability of such cooperative equilibria is essential for designing robust multi-agent protocols for distributed systems, cyber defense, or web-scale coordination tasks.

Our main result provides a sufficient condition for full cooperation in distributed PD games, originally established by \citet{monderer1999distributed}, to environments with \emph{network-delayed communication}. We derive a generalized equilibrium condition that explicitly accounts for the propagation delays induced by the communication network, thereby characterizing when cooperation can be sustained under limited and delayed information exchange. Building on this characterization, we analyze several extreme regimes and illustrate how the \emph{number of locations} and the \emph{network diameter} jointly determine the feasibility of cooperation. 
Our results offer insights into the design and analysis of multi-agent systems operating over networks, highlighting how communication latency shapes the emergence and sustainability of cooperative behavior in distributed environments.

\section{Distributed Games with Network Delays}
\label{sec:model}

We extend the distributed games framework of \citet{monderer1999distributed} 
by constraining inter-agent communication to a network topology. 
We assume messages are not globally broadcast but instead propagate along a weighted graph 
connecting locations. This induces location-dependent information sets and 
changes the equilibrium analysis.
Consider a finite set of players $F=\{1,\dots,m\}$ and a finite set of locations 
$L=\{1,\dots,n\}$. At each location $i \in L$ there is a one-shot strategic-form game
\[
G_i = \big( (S_{ij})_{j\in F}, (u_{ij})_{j\in F} \big),
\]
where $S_{ij}$ is the finite action set of player $j$'s agent at location $i$, 
and $u_{ij} : \times_{k\in F} S_{ik} \to \mathbb{R}$ is $j$'s payoff at $i$.
The game proceeds over $n$ rounds. At each round exactly one location is active and 
its local game is played. The order is chosen at the start of play according to a 
distribution $\lambda$ over permutations $\pi$ of $L$. For a specific $\pi$, let 
$i_r := \pi(r)$ denote the active location at round $r$.

Communication occurs over a fixed, connected, undirected, weighted graph 
$\mathcal{N}=(V,E,\delta)$ where $V=L$. An edge $(i,i')\in E$ means that agents at 
locations $i$ and $i'$ can directly exchange messages. Each edge is associated with 
a non negative integer delay $\delta(i,i') \geq 0$, representing the number of rounds 
required for a message to traverse the edge. In the distributed games model of \citet{monderer1999distributed}, $\delta(i,i') = 0$ for every edge $(i,i')$.  
For any two locations $i,i' \in L$, the \emph{propagation distance} is defined as follows:
\[
d_{\mathcal{N}}(i,i') \;=\; \min \Big\{ \sum_{\ell=0}^{k-1} \delta(i_\ell,i_{\ell+1}) : 
(i=i_0,i_1,\dots,i_k=i') \text{ is a path in } \mathcal{N} \Big\}.
\]
Thus $d_{\mathcal{N}}(i,i')$ is the minimum number of rounds needed for a message 
sent at location $i$ to reach location $i'$, assuming immediate relaying at each 
intermediate node.
Each player $j$ has a finite message alphabet $M_j$ with a null symbol $\varphi$.
At round $r$ with active location $i_r$:
\begin{enumerate}
  \item \textbf{Action stage:} Agents at $i_r$ simultaneously choose actions 
  $a_{i_r j} \in S_{i_r j}$.
  \item \textbf{Message stage:} Each agent sends messages to its neighbors 
  $N(i_r)$ in $\mathcal{N}$. A message sent from $i$ to $i'$ via edge $(i,i')$ 
  becomes available to $i'$ only after $\delta(i,i')$ further rounds. If relayed 
  through a path, the total delay is the path length in the weighted sense, 
  i.e.\ $d_{\mathcal{N}}(i,i')$.
\end{enumerate}

\paragraph{Information sets}
The formulation of the information sets is identical to the that in \citet{monderer1999distributed}. Each agent $i_j$ maintains the set of messages it has received so far. 
At the start of round $r$, denote this set by $\mathcal{M}^{(j)}_{i}(r)$. 
The signal of agent $i_j$ at the action stage of round $r$ is
\[
Z^{(a)}_{i_j}(r) = \big(i_r, \mathcal{M}^{(j)}_{i}(r)\big),
\]
and at the message stage it is
\[
Z^{(m)}_{ij}(r) = \big(i_r, a_{i_r}, \mathcal{M}^{(j)}_{i}(r)\big).
\]
Two histories lie in the same information set for $i_j$ if the corresponding $Z$ 
tuples coincide. 

\paragraph{Strategies and payoffs}
A behavioral strategy for player $j$ is a pair $\sigma_j=(f_j,g_j)$ where
\[
f_{i_j}: Z^{(a)}_{i_j} \to S_{ij}, 
\qquad 
g_{i_j}: Z^{(m)}_{i_j} \to M_j^{i}.
\]
The expected payoff for player $j$ under strategy profile $\sigma$ and ordering 
$\pi \sim \lambda$ is
\[
u_j(\sigma) =  \sum_{i \in L} E_{i_j}(\sigma),
\]
where $\sigma$ is a strategy profile and $E_{i_j}(\sigma)$ is the expected payoff of player $j$ from location $i$, when all players use the strategy profile $\sigma$.

\subsection{Equilibrium}

We define an equilibrium as a strategy profile $\sigma$ such that no player~$j$ 
can increase her total payoff~$u_j$ by deviating unilaterally, given the communication constraints 
and propagation delays determined by $d_{\mathcal{N}}(\cdot,\cdot)$.
To analyze equilibrium feasibility under delayed communication, we first formalize
how long it takes for information about a deviation to become observable across the network. In this context, an \emph{alarm} refers to the message triggered by an agent upon observing a deviation from cooperation. This message then propagates through the network according to the communication delays.

\begin{lemma}[Propagation–Detectability Bound]
\label{lem:detectability}
Fix a realized order~$\pi$ and consider a deviation that occurs at location~$i$
during the \emph{action stage} of round~$t$.
If, in the \emph{message stage} of the same round,
agents immediately relay an alarm along all outgoing edges, 
then for any other location~$i' \in L$ the earliest round
in which that alarm can reach~$i'$ is
\[
t + d_{\mathcal{N}}(i,i'),
\]
where $d_{\mathcal{N}}(i,i')$ denotes the weighted shortest-path delay in the communication network~$\mathcal{N}$.
In particular, no agent at~$i'$ can condition its action on the deviation before round~$t + d_{\mathcal{N}}(i,i')$.
\end{lemma}

\begin{proof}
By definition $d_{\mathcal{N}}(i,i')$ is the total delay of a message propagate from $i$ to $i'$. Since messages propagate synchronously and can only be forwarded one hop per round, the minimal time required for an alarm from $i$ to reach $i'$ is exactly $d_{\mathcal{N}}(i,i')$ rounds.
Therefore, no message originating at $t$ can be available at $i'$ before round $t + d_{\mathcal{N}}(i,i')$,
which establishes the bound.
\end{proof}

\section{Prisoner’s Dilemma with Network Delays}

We now focus on the parallel Prisoner’s Dilemma (PD) setting, where the following game is played at each location $i \in L$:

\[
\begin{array}{c|cc}
    & D & C \\
\hline
D & (a,a) & (b,0) \\
C & (0,b) & (c,c) \\
\end{array}
\]

with payoffs satisfying $b > c > a > 0$.
Messages propagate on the weighted network $\mathcal{N}=(L,E,\delta)$, and let the diameter be
\[
\tau \;=\; \mathrm{Diam}(\mathcal{N}) \;=\; \max_{i,i'\in L} d_{\mathcal{N}}(i,i').
\]
Thus, any alarm raised at some location reaches all locations no later than $\tau$ rounds after it was sent. As in \citet{monderer1999distributed}, we assume a uniform distribution of locations' order.

\paragraph{Strategy (network with relay).}
Our goal now is to characterize whenever an equilibrium in which both agents cooperate in all locations exist.
By similar arguments as in \citet{monderer1999distributed}, it is sufficient to consider strategies of the following form.
\begin{enumerate}
  \item \emph{Cooperate:} In the action stage at every active location, each agent plays $C$ unless it has set an alarm flag.
  \item \emph{Alarm and relay:} If any agent observes a deviation from $C$ at the currently active location (either by its own opponent or via a received message about a past deviation) it sets an alarm and, in the message stage, sends an alarm to all neighbors. Upon receiving any alarm, an agent sets its alarm permanently and relays it in subsequent message stages.
  \item \emph{Punish:} If an agent’s alarm is set before it acts at some future location, it plays $D$ there (against the identified deviator if identification is feasible; otherwise symmetric $D$).
\end{enumerate}

By the propagation–detectability bound (Lemma \ref{lem:detectability}), a deviation in round $t$ can affect play no later than round $t+\tau$.
If a player deviates in the action stage of round $t$, under a delay $\tau$, the deviation payoff (i.e., the maximal payoff since we assume a delay of $\tau$) is
\[
v_t^{\tau} \;=\; (t-1)c \;+\; \min\{\tau+1,\,n-t+1\}\,b \;+\; (n-t-\tau)_+\,a ,
\]

where $(x)_+ \coloneq \max\{x,0\}$.
The term $\min\{\tau+1,n-t+1\}b$ constitutes the main modification relative to Eq. (3.1) in \citet{monderer1999distributed}. If the diameter is sufficiently small (roughly $\tau << n)$ to ensure that the alarm message reaches some of the agents of the second player during the game, then the maximum number of locations in which the deviating player can play with its agents $D$ before the agents of the second player receive the alarm is $\tau+1$. Conversely, if the diameter is too large ($\tau > \alpha n, \alpha>1$), the game may terminate before the alarm reaches the second player. In that case the controlling term becomes $n-t+1$. For this reason, we also adjust the third term in Eq. (3.1) of \citet{monderer1999distributed} to $(n-t-\tau)_+a$, because the strategy profile $(D,D)$ is not played.

We define the \emph{average deviation payoff} as the mean payoff a deviating player obtains per stage when deviating once and the alarm propagates within $\tau$ rounds. Formally,

\[
v^{1,\tau} \;=\; \frac{1}{n}\sum_{t=1}^n v_t^{\tau}.
\]
The equilibrium requirement (no profitable deviation on average) is then given by
$
v^{1,\tau} \;\le\; n c
$.
Evaluating the sums (split at $t=n-\tau$) and simplifying yields the following explicit bound: 

\begin{theorem}
\label{thrm:cooperation}
Full cooperation in equilibrium is sustainable if
\[
b \;\le\; c \;+\; (c-a)\cdot \frac{(n-\tau)(\,n-\tau-1\,)}{(2n-\tau)(\tau+1)}.
\tag{EC$_\tau$}
\]    
\end{theorem}

\begin{proof}    
We begin with the deviation payoff for a deviation of length $\tau$ at stage $t$:
\[
v_t^\tau = (t-1)c + \min\{\tau+1, n-t+1\} b + (n-t-\tau)_+ a.
\]

The average deviation payoff is:
\[
v^{1,\tau} = \frac{1}{n}\sum_{t=1}^{n} v_t^\tau.
\]

The equilibrium condition requires $v^{1,\tau} \le nc$.
Expanding the sum:
\[
\sum_{t=1}^{n} v_t^\tau = 
c\sum_{t=1}^{n}(t-1) +
b\sum_{t=1}^{n}\min\{\tau+1,n-t+1\} +
a\sum_{t=1}^{n}(n-t-\tau)_+.
\]

Compute each part:
\begin{align*}
\sum_{t=1}^{n}(t-1) &= \frac{n(n-1)}{2},\\
\sum_{t=1}^{n}\min\{\tau+1,n-t+1\} &= (n-\tau)(\tau+1)+\frac{\tau(\tau+1)}{2} = (\tau+1)\Big(n-\frac{\tau}{2}\Big),\\
\sum_{t=1}^{n}(n-t-\tau)_+ &= \frac{(n-\tau-1)(n-\tau)}{2}.
\end{align*}

Thus:
\[
\sum_{t=1}^{n} v_t^\tau = 
c\frac{n(n-1)}{2} +
b(\tau+1)\Big(n-\frac{\tau}{2}\Big) +
a\frac{(n-\tau-1)(n-\tau)}{2}.
\]

Next, we average payoffs and get:
\[
v^{1,\tau} = \frac{1}{n}\Big[
c\frac{n(n-1)}{2} +
b(\tau+1)\Big(n-\frac{\tau}{2}\Big) +
a\frac{(n-\tau-1)(n-\tau)}{2}
\Big].
\]

Applying the equilibrium condition yields:
\[
\frac{1}{n}\Big[
c\frac{n(n-1)}{2} +
b(\tau+1)\Big(n-\frac{\tau}{2}\Big) +
a\frac{(n-\tau-1)(n-\tau)}{2}
\Big] \le nc.
\]

Multiply by $n$:
\[
c\frac{n(n-1)}{2} +
b(\tau+1)\Big(n-\frac{\tau}{2}\Big) +
a\frac{(n-\tau-1)(n-\tau)}{2}
\le n^2c.
\]

Isolating $b$:
\begin{align*}
b(\tau+1)\Big(n-\frac{\tau}{2}\Big) &\le \frac{c(n^2+n) - a(n-\tau-1)(n-\tau)}{2},\\
b &\le \frac{c n(n+1) - a(n-\tau-1)(n-\tau)}{(2n-\tau)(\tau+1)}.
\end{align*}

Factoring terms to obtain (EC$_\tau$) concludes the proof.



\end{proof}

Notice that the fraction
\[
\frac{(n-\tau)(n-\tau-1)}{(2n-\tau)(\tau+1)}
\]
monotonically decreases in \(\tau\) (for fixed \(n\)), reflecting that larger propagation delays reduce the effective future punishment window available to deter deviations; when \(\tau\) approaches \(n\) the RHS collapses, making cooperation harder, or even impossible.

\subsection{Asymptotic Case Analysis}

Using the cooperation equilibrium result stated above, we now move on to analyzing several key cases with concrete design implications on distributed communication. In particular, we consider several cases of the relationship between the number of locations $n$ and the communication network diameter $\tau$, and study the asymptotic behavior of (EC$_\tau$) for each configuration.

\paragraph{No delay}
If \(\tau=0\) then the propagation is effectively instantaneous, and (EC$_\tau$) reduces to the condition of \citet{monderer1999distributed}:
\[
b \le c + \frac{n-1}{2}\,(c-a).
\]

\paragraph{Full delay}
If \(\tau \ge n-1\) then \(n-\tau \le 1\), hence the numerator \((n-\tau)(n-\tau-1)\) in (EC$_\tau$) is zero and the right-hand side collapses to \(c\). Thus (EC$_\tau\)) becomes $b \le c$,
which cannot hold in the Prisoner’s Dilemma (where \(b>c\)). Intuitively, when alarms cannot reach all locations before the horizon ends, network-wide punishment is infeasible and cooperation unravels.




\paragraph{Scale-free communication network.}
In many web and online systems, the underlying communication structure follows a \emph{scale-free} topology, characterized by a few highly connected hubs and many low-degree peripheral nodes \citep{barabasi2000scale}. Such networks are known to exhibit very short typical path lengths: both theoretical and empirical analyses show that their diameter grows sub-linearly with network size \cite{chung2002average,cohen2003scale}. Hence, $\tau = o(n)$, meaning that communication delay grows much slower than the number of locations. Substituting this scaling into (EC$_\tau$) yields 
\[
\frac{(n-\tau)(n-\tau-1)}{(2n-\tau)(\tau+1)} \sim \frac{n}{2(\tau+1)}.
\]
Therefore, the RHS grows roughly like
\[
c + (c-a)\,\frac{n}{2(\tau+1)},
\]
showing that with many locations (large \(n\)) cooperation becomes easier to sustain provided \(\tau\) is fixed and relatively small compared to \(n\).
That is, in scale-free communication networks, cooperation becomes increasingly sustainable as the system expands: despite growing size, the high connectivity of hubs ensures that deviations are detected and punished quickly, effectively maintaining a large "shadow of the future" across the network.

\paragraph{Proportional delay}
Consider the case of \(\tau=\alpha n\) with some fixed \(0<\alpha<1\). Taking \(n\) large and keeping \(\alpha\) fixed yields
\[
\lim_{n\to\infty}\frac{(n-\tau)(n-\tau-1)}{(2n-\tau)(\tau+1)}
=
\frac{(1-\alpha)^2}{\alpha(2-\alpha)}.
\]
Thus (EC$_\tau$) becomes, asymptotically,
\[
b \;\le\; c \;+\; (c-a)\,\frac{(1-\alpha)^2}{\alpha(2-\alpha)}.
\]
This shows a smooth interpolation: when \(\alpha\) is small (delays negligible relative to horizon) the bound is generous; as \(\alpha\to 1\) (delays comparable to the horizon), the right-hand side tends to \(c\), eliminating incentives for cooperation.
To sum up:
\begin{itemize}
  \item If \(\tau\) is \emph{much smaller} than \(n\) (fast propagation relative to horizon), network effects are minor and cooperation is readily supported for large \(n\).
  \item If \(\tau\) is \emph{comparable} to \(n\), the punishment window shrinks and cooperation becomes fragile; in the extreme \(\tau \ge n-1\) it is impossible in the averaged sense.

\end{itemize}

These findings highlight and quantify two main levers for preserving cooperation: extending the effective horizon \(n\) (through more interactions or parallel locations) and reducing propagation delays \(\delta(\cdot,\cdot)\) (via network design or faster communication).


\section{Discussion}

We analyzed cooperation in distributed games under network-constrained communication, deriving an equilibrium condition that links cooperation feasibility to the network diameter~$\tau$ and the number of locations~$n$. Our results show that communication delays shrink the effective punishment horizon, making cooperation harder as~$\tau$ increases but easier as~$n$ grows. In scale-free networks, cooperation remains sustainable even at large scales, as highly connected hubs enable rapid detection of deviations. Our delayed communication model also contributes to the study of decision making with broadcast communication in multi-agent systems \citep{bahar2005sequential}.

A promising future direction is to establish a folk-theorem–like characterization of feasible equilibrium outcomes in general distributed games \cite{mailath2006repeated}, which could provide a general framework for studying strategic behavior in distributed multi-agent environments.
Another potential future direction is to move beyond our high-level, worst-case analysis based solely on the network diameter~$\tau$, and develop a finer characterization that accounts for the exact network topology. In real systems, agents occupy positions with varying centrality: those at highly connected nodes can transmit and receive information rapidly, while peripheral agents experience longer delays and weaker enforcement of cooperative norms. Understanding how such structural heterogeneity affects equilibrium incentives could yield a richer theory of distributed cooperation, bridging network topology with strategic stability.

\end{document}